\documentclass[pra,twocolumn,superscriptaddress,10pt,noshowpacs]{revtex4}
\usepackage[english]{babel}
\usepackage[T1]{fontenc}
\usepackage[utf8]{inputenc}
\usepackage{graphicx,epstopdf}
\usepackage{amssymb}
\usepackage{amsmath}

\usepackage{amsfonts}
\usepackage{bbm}
\usepackage{color}
\usepackage{latexsym}
\usepackage{caption}
\usepackage{subcaption}
\usepackage{times,txfonts}

\newcommand{\beq}{\begin{equation}}
\newcommand{\eeq}{\end{equation}}
\newcommand{\bea}{\begin{eqnarray}}
\newcommand{\eea}{\end{eqnarray}}
\begin{document}

\title{Massless fermions in black string spacetime}

\author{R. Darlla}
\affiliation{Universidade Federal do Cear\'a, Departamento de F\'{i}sica, 60455-760, Fortaleza, CE, Brazil}

\author{\"{O}. Ye\c{s}ilta\c{s}}
\affiliation{Department of Physics, Faculty of Science, Gazi University, 06500 Ankara, Turkey}
\email{yesiltas@gazi.edu.tr}

\author{J. Furtado}
\email{job.furtado@ufca.edu.br}
\affiliation{Universidade Federal do Cariri, Centro de Ci\^encias e Tecnologia, 63048-080, Juazeiro do Norte, CE, Brazil}
\affiliation{Department of Physics, Faculty of Science, Gazi University, 06500 Ankara, Turkey}

\date{\today}

\begin{abstract}

In this paper we investigate the behaviour of massless fermions in the black string spacetime by computing the eigenvalues and eigenfunctions of the Weyl equations. These solutions allowed us to study the behaviour of such massless fermions in terms of the cosmological constant, the black string's mass and the radial distance of the particle from the black string. The solutions, written in terms of Parabolic Cylinder functions and Laguerre polynomials, were obtained for a particle far from the black string and around the horizon event. For the particle around the event horizon, for all configuration of parameters, the energy eigenvalues are complex-valued, indicating QNM similarly to the case of spherical black holes. For the particle far from the black string, the energies derived from the Weyl equation set up conditions on the parameters in order to keep the energy as a real valued parameter. 

\end{abstract}

\maketitle

\section{Introduction}


Black holes are obtained as solutions of Eintein's equations and play a very relevant role in Physics, since such objects can be used to understand how space-time is established after a gravitational collapse. Although there is a natural tendency to study spherically symmetric black holes, especially in space-times with a vanishing cosmological constant, the study of such objects with different topologies also become something of interest. Space-times with a negative cosmological constant are the backgroung for the existence of black holes with cylindrical symmetry. A cylindrically symmetric black hole solution, namely black string, for a four-dimensional Einstein-Hilbert action was proposed in \cite{Lemos:1994xp}, in the context of the classical theory of gravitation. These solutions have gained a great deal of relevance in recent years due to technological evolution, as evidenced by the detection of gravitational waves by LIGO and VIRGO and the first images of supermassive black holes \cite{EventHorizonTelescope:2022wkp, EventHorizonTelescope:2019dse, LIGOScientific:2016aoc, LIGOScientific:2017vwq}.


Solutions with cylindrical symmetry are particularly important for cosmology, as in the study of the evolution of the universe. In the phase transitions that may have occurred after the big bang, so-called topologically stable defects are studied, such as cosmic strings; these have generated very interesting results, such as density fluctuation, which explains the creation of galaxies \cite{Vilenkin:1984ib}. A black string is such kind of solution. In order to be more accurate, black strings are higher-dimensional solutions of the Einstein equations in D-dimensional spacetime \cite{Duff:1987cs}. These vacuum objects can be understood as symmetric generalizations of black holes under translations, with the event horizon topologically equivalent to $S_2 \times R$ (or $S_2\times S_1$ in the case of black rings), and spacetime is asymptotically $ M_{D-1} \times S_{1} $ for a zero cosmological constant or $   AdS_{D-1} \times S_{1} $ for a negative one \cite{Emparan:2001wn,Bellucci:2010gb}. A black string is a particular case of a black p-brane solution (with $p = 1$), where $p$ is the number of extra spatial dimensions to ordinary space. Hence, a p-brane gives rise to a (p + 1)-dimensional world-volume in spacetime. In string theory, a black string is described by a D1-brane surrounded by a horizon. It is widely known that since an open string propagates through spacetime, its endpoints must lie on a D2-brane on which it satisfies the Dirichlet boundary condition, and the dynamics on the D-brane world-volume is described by a gauge field theory \cite{Polchinski:1995mt, Polchinski:1994fq}. In some models, our universe is understood as a higher-dimensional world-brane; thus, the gravitational collapse of galactic matter would produce black holes lying on the brane. Black string solutions are nevertheless unstable to long-wavelength perturbations (at least in asymptotically flat spaces) since the localized black hole is entropically preferred to a long segment of string. The string’s horizon therefore has a tendency to form a line of black holes. However, in AdS, the space acts as a confining box which prevents fluctuations of long wavelengths from developing. For instance, in a Randall–Sundrum domain wall in a five-dimensional AdS spacetime, the black string is stable at least far from the AdS horizon. Near the horizon, the string is likely to pinch off to become a stable shorter cigar-like singularity \cite{Chamblin:1999by}.

More recently, there are several works in the literature investigating black strings in different scenarios. In \cite{Darlla:2023qgf}, the authors investigate the behaviour of black string thermodynamics and the orbits of massive and massless particles around black strings in the context of Rainbow Gravity, which is an effective way of treating quantum gravity. In \cite{Lima:2022pvc}, a bounce between a black string and a traversable wormhole is investigated. The authors of \cite{Carvalho:2022eli} investigated the fractalization of the black string's event horizon in the context of unparticles gravity and in \cite{Muniz:2022otq} the authors studied the static and rotating black string's thermodynamics in the context of a bilocal gravity. 

In this paper we investigate the behaviour of massless fermions in the black string spacetime by computing the eigenvalues and eigenfunctions of the Weyl equations. These solutions allowed us to study the behaviour of such massless fermions in terms of the cosmological constant, the black string's mass and the radial distance of the particle from the black string. This paper is organized as follows: in the next section we present a brief review of the black string spacetime, compute the Weyl equation for the two spinor components and decoupled them into two Klein-Gordon-like equations for each spinor component. In section III we obtained the solutions for a particle far from the black string and around the horizon event. In section IV we highlight our conclusions.

\section{Black string spacetime}

First, let's review the black string solution for Einstein's equations with negative cosmological constant obtained by Lemos in \cite{Lemos:1994xp}, within the scenario of the classical theory of gravitation. Let's consider the following line element
\begin{equation}\label{3}
    ds^2=-A(r)dt^2+\frac{dr^2}{A(r)}+r^2d\phi^2+\alpha^2r^2dz^2,
\end{equation}
where $-\infty < t < \infty$ and $0 \leq r < \infty$, $0 \leq \phi \leq 2 \pi$,$-\infty < z < \infty$ define the radial, angular and axial coordinates, respectively. The parameter $\alpha$ is such that $\alpha^2 \equiv -\Lambda/3 > 0$, where $\Lambda$ is the cosmological constant. For the black string the Einstein-Hilbert effective action requires the cosmological constant contribution, so that,

\begin{eqnarray}\label{24}
    S_u = \frac{1}{2\kappa^2} \int d^4x \sqrt{-g} \left(R-2\Lambda\right).
\end{eqnarray}
where $\kappa=8\pi G$ and $R$ is the Ricci scalar. The function $A(r)$ is given by
\begin{equation}
    A(r)=\alpha^2r^2-\frac{4\mu}{\alpha r},
\end{equation}
where $\mu$ is identified as the black string's mass. In order to compute the Dirac equation in the black string spacetime, it is suitable to choose the orthonormal 1-forms which satisfies $g_{\alpha\beta}=\eta_{AB}e^A_{\alpha}e^B_{\beta}$. From the metric in (\ref{3}) and the function $A(r)$ we can calculate the non-vanishing christoffel as
\begin{eqnarray}
    \Gamma^0_{01}&=&\Gamma^0_{10}=\frac{r^3\alpha^3+2\mu}{r^4\alpha^3-4\mu r},\\
    \Gamma^1_{00}&=&r^3\alpha^4-2\alpha\mu-\frac{8\mu^2}{r^3\alpha^2},\\
    \Gamma^1_{11}&=&-\frac{r^3\alpha^3+2\mu}{r^4\alpha^3-4r\mu},\\
    \Gamma^1_{22}&=&-r^3\alpha^2+\frac{4\mu}{\alpha},\\
    \Gamma^1_{33}&=&-r^3\alpha^4+4\alpha\mu,\\
    \Gamma^2_{12}&=&\Gamma^2_{21}=\frac{1}{r},\\
    \Gamma^3_{13}&=&\Gamma^2_{31}=\frac{1}{r}.
\end{eqnarray}
Hence, the connection $\omega_{AB\mu}$ are computed from
\begin{equation}
    \omega_{AB\mu}=g_{\alpha\beta}e^\alpha_A\left(\partial_\mu e_B^\beta+\Gamma^\beta_{\mu\lambda}e_B^\lambda\right).
\end{equation}
Let us summarize all the non-vanishing contributions of $\omega_{AB\mu}$, which are:
\begin{eqnarray}
    \omega_{01t}&=&-\alpha^2 r-\frac{2\mu}{\alpha r^2}\\
    \omega_{12\phi}&=&-\sqrt{\alpha^2 r^2-\frac{4\mu}{\alpha r}}\\
    \omega_{13z}&=&-\alpha\sqrt{\alpha^2 r^2-\frac{4\mu}{\alpha r}}.
\end{eqnarray}

At this point we are able to compute the spin connection (Fock-Ivanenko coefficients) from
\begin{equation}
    \Gamma_\mu=\frac{1}{4}\omega_{AB\mu}\gamma^A\gamma^B.
\end{equation}
All four contributions are the following:
\begin{eqnarray}
    \Gamma_0&=&\frac{1}{4}\left(-\alpha^2 r-\frac{2\mu}{\alpha r^2}\right)\gamma^0\gamma^1,\\
    \Gamma_1&=&0,\\
    \Gamma_2&=&\frac{1}{4}\left(-\sqrt{\alpha^2 r^2-\frac{4\mu}{\alpha r}}\right)\gamma^1\gamma^2,\\
    \Gamma_3&=&\frac{1}{4}\left(-\alpha\sqrt{\alpha^2 r^2-\frac{4\mu}{\alpha r}}\right)\gamma^1\gamma^3.
\end{eqnarray}

We can finally compute the massless Dirac equation (Weyl equation) directly from
\begin{equation}
    i\bar{\gamma^{\mu}}D_{\mu}\psi=0,
\end{equation}
where $D_{\mu}\psi\equiv(\partial_{\mu}+\Gamma_{\mu})\psi$ and the curved spacetime Dirac matrices $\bar{\gamma^{\mu}}$ are related with the flat spacetime Dirac matrices by $\bar{\gamma^{\mu}}=e_A^{\mu}\gamma^A$. Since we are interested in massless fermions, let us consider the following ansatz for the spinor
\begin{eqnarray}
    \psi=e^{-iEt-im\phi-ikz}\left(\begin{array}{cc}
         \psi_1  \\
         \psi_2 
    \end{array}\right),
\end{eqnarray}
where $E$ stands for the fermions energy, $m$ is the orbital angular momentum and $k$ is a quantum number related to the fact that the system exhibit symmetry along the z axis. In our case the Dirac matrices are such that $\gamma^0=\hat{I}_2$, being $\hat{I}_2$ the order two identity matrix and $\gamma^i=\sigma^i$, where $\sigma^i$ are the Pauli matrices. In this representation we are able the write the Weyl equation in matricial form as follows:
\begin{eqnarray}
    \left(\begin{array}{cc}
     \frac{E}{\sqrt{A(r)}}+\frac{k}{\alpha r}    & \Delta_1 \\
      \Delta_2   & \frac{E}{\sqrt{A(r)}}-\frac{k}{\alpha r}
    \end{array}\right)\left(\begin{array}{cc}
         \psi_1  \\
         \psi_2 
    \end{array}\right)=0.
\end{eqnarray}
The off-diagonal terms are writen as
\begin{eqnarray}
    \nonumber\Delta_1&=&i\sqrt{A(r)}\partial_r-\frac{i}{4\sqrt{A(r)}}\left(\alpha^2 r+\frac{2\mu}{\alpha r^2}\right)-\frac{i\sqrt{A(r)}}{2r}-\frac{m}{r},\\\\
    \nonumber\Delta_2&=&i\sqrt{A(r)}\partial_r-\frac{i}{4\sqrt{A(r)}}\left(\alpha^2 r+\frac{2\mu}{\alpha r^2}\right)-\frac{i\sqrt{A(r)}}{2r}+\frac{m}{r}.\\
\end{eqnarray}
Notice that we have two coupled first-order partial differential equations for $\psi_1$ and $\psi_2$. For the sake of clarity, we will define the functions $\Sigma_1(r)$, $\Sigma_2(r)$ and $\Lambda(r)$ as
\begin{eqnarray}
    \Lambda(r)&=&\frac{i}{4\sqrt{A(r)}}\left(\alpha^2 r+\frac{2\mu}{\alpha r^2}+\frac{2A(r)}{r}\right),\\
    \Sigma_1(r)&=&\frac{E}{\sqrt{A(r)}}+\frac{k}{\alpha r},\\
    \Sigma_2(r)&=&\frac{E}{\sqrt{A(r)}}-\frac{k}{\alpha r},
\end{eqnarray}
so that we can write the equations for $\psi_1$ and $\psi_2$ as
\begin{eqnarray}\label{a1}
    \Sigma_1(r)\psi_1(r)&=&-i\sqrt{A(r)}\partial_r\psi_2(r)+\Lambda(r)\psi_2(r)+\frac{m}{r}\psi_2(r)\\
    \Sigma_2(r)\psi_2(r)&=&-i\sqrt{A(r)}\partial_r\psi_1(r)+\Lambda(r)\psi_1(r)-\frac{m}{r}\psi_1(r).
\end{eqnarray}

Multiplying (\ref{a1}) by $\sqrt{A(r)}$ and arranging the equation, we obtain 

\begin{eqnarray}
  E \psi_1 &=&-i A(r)\psi'_{2}+\left(\frac{i}{4}S(r)+\frac{(m-k/\alpha)\sqrt{A(r)}}{r}\right)\psi_{2} \\
  E \psi_2 &=&-i A(r)\psi'_{1}+\left(\frac{i}{4}S(r)-\frac{(m-k/\alpha)\sqrt{A(r)}}{r}\right)\psi_{1} 
\end{eqnarray}
where $S(r)=\alpha^{2}r+\frac{2\mu}{\alpha r^{2}}+\frac{2A(r)}{r}$.
Hence, we obtain the second order equations as below
\begin{eqnarray}\label{d1}
   \nonumber-A(r)^{2}\psi''_{1}+\left(\frac{1}{2}A(r)S(r)-A(r)A'(r)\right)\psi'_{1}+V_1(r)\psi_{1}&=&  E^{2}\psi_1\\\\
   \nonumber-A(r)^{2}\psi''_{2}+\left(\frac{1}{2}A(r)S(r)-A(r)A'(r)\right)\psi'_{1}+V_2(r)\psi_{1}&=&  E^{2}\psi_2\\ \label{d01}
\end{eqnarray}
where $S(r)=\alpha^{2}r+\frac{2\mu}{\alpha r^{2}}+\frac{2A(r)}{r}$ and
   
   \begin{eqnarray}\label{d2}
     \nonumber V_1(r) &=& -\frac{(k-m\alpha)^{2}}{r^{2}\alpha^{2}}A(r)-\frac{i}{r^{2}}A(r)^{3/2}(m-k/\alpha)-\frac{S(r)^{2}}{16}\\
     &&+\frac{i(m-k/\alpha)}{2r}A'\sqrt{A}+\frac{1}{4}A(r)S'(r) \\
     \nonumber V_2(r) &=& -\frac{(k-m\alpha)^{2}}{r^{2}\alpha^{2}}A(r)+\frac{i}{r^{2}}A(r)^{3/2}(m-k/\alpha)-\frac{S(r)^{2}}{16}\\
     &&-\frac{i(m-k/\alpha)}{2r}A'\sqrt{A}+\frac{1}{4}A(r)S'(r). \label{d02}
   \end{eqnarray}

Now we are able to seek for the eigenfunctions and eigenvalues of the system. The solutions will be addressed in the next section.

\section{Solutions}

In order to obtain the solutions for our system, let us use a transformation for the wavefunction $\psi_1(r)$ in (\ref{d1}) as given below
\begin{equation}\label{d22}
  \psi_1(r)= \exp\left(\int^{r} \frac{r'^{3}\alpha^{3}+2\mu+2r'\alpha A(r')-2r'^{2}\alpha A'(r')}{4r'^{2}\alpha A(r')} dr' \right)Y_1(r)
\end{equation}
and dividing (\ref{d1}) by $-A(r)^{2}$, we obtain
\begin{equation}\label{d3}
   Y''_{1}(r) +U(r)Y_1(r)=0
\end{equation}
where
\begin{eqnarray}\label{d4}
\nonumber U(r)&=&\frac{E^{2}}{A(r)^{2}}+\frac{k^{2}-2km\alpha+m^{2}\alpha^{2}}{\alpha^{2}r^{2}A(r)^{2}}+\frac{i(-k+m\alpha)}{r^{2}\alpha A(r)} \\ 
  &&-\frac{i(-k+m\alpha)A'(r)}{2r\alpha A(r)^{3/2}}+\frac{A'(r)^{2}}{A(r)^{2}}-\frac{A''(r)}{2A(r)}.
\end{eqnarray}
As it is seen from (\ref{d3}) and (\ref{d4}), the system is not exactly solvable. We may mention the behaviour of the $U(r)$ function at infinity by expanding $U(r)$ in binomial series around $r=r_h$, which is the horizon radius, and $r=\infty$. \\

\subsection{Solution for $r=r_h$}
In this case we look at the behaviour of $U(r)$ for the first three terms,i.e;
\begin{equation}\label{d5}
   U_{a}(r)=C_1+C_2(r-r_h)+C_3 (r-r_h)^{2}+O(r-r_h)^{3}
\end{equation}
where $C_1$, $C_2$, and $C_3$ are the complex functions depending on the parameters $a$, $\alpha$, $\mu$, $m$, and $k$. Since the constants $C_1$, $C_2$, and $C_3$ are too lengthy, we will omit them for the sake of simplicity and clarity. Thus, we obtain

\begin{eqnarray}\label{d6}
   \nonumber Y_1(r)&=&c_1 D_{\nu}\left(\gamma+(-1)^{1/4}\sqrt{2}C^{1/4}_{3}r\right)\\
   &&+c_2 D_{\upsilon}\left(i\gamma+i(-1)^{1/4}\sqrt{2}C^{1/4}_{3}r\right)
\end{eqnarray}
where

\begin{eqnarray}
  \nu &=& \frac{iC^{2}_{2}-4iC_{1}C_{3}-4C^{3/2}_{3}}{8C^{3/2}_{3}} \\
  \gamma &=& -\frac{(-1)^{1/4}(-C_2+2aC_3)}{\sqrt{2}C^{3/4}_{3}} \\
  \upsilon &=& \frac{-iC^{2}_{2}+4iC_{1}C_{3}-4C^{3/2}_{3}}{8C^{3/2}_{3}}
\end{eqnarray}

The two terms in your solution represent linearly independent solutions to the differential equation, with $c_1$ and $c_2$ acting as integration constants and $D_{p}(b r)$ stand for the  Parabolic Cylinder functions. 
\begin{figure}
    \centering
    \includegraphics[scale=0.5]{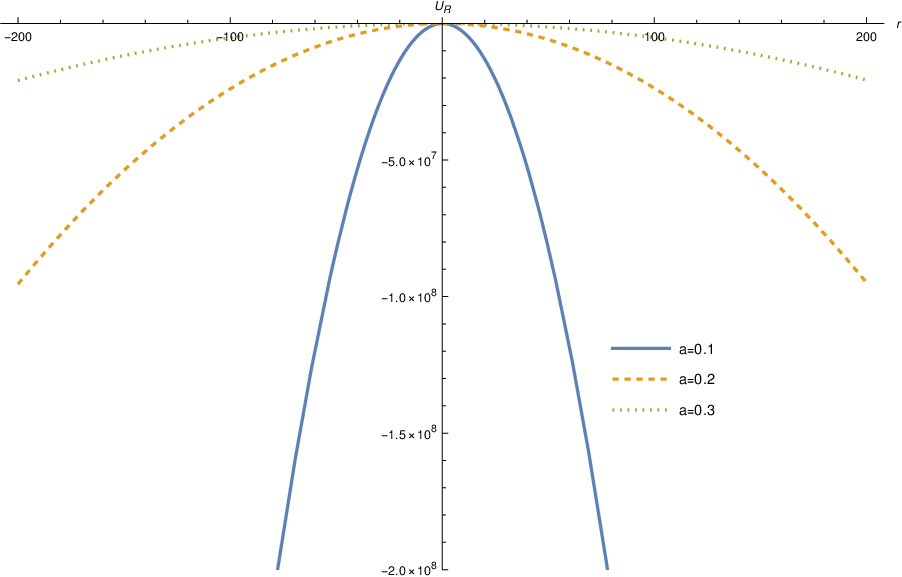}
    \caption{The graph of the real part of the potential $U_a(r)$ in (\ref{d4}) $e = 10; \alpha = 0.5; m = 1, k = 2, \mu = 1 $. And $a=0.1, 0.2, 0.3$ for the blue, yellow and green curves respectively. }
    \label{fig:enter-label}
\end{figure}
If we continue with (\ref{d01}), similar steps can be followed. It is important to highlight here that (\ref{d02}) can be obtained directly from (\ref{d2}) by applying the following interchange, $m \rightarrow -m$ and $k \rightarrow -k$, so that we obtain the solutions $Y_2(r)$ as given by

\begin{eqnarray}\label{d7}
   \nonumber Y_2(r)&=&c_3 D_{\bar{\nu}}\left(\bar{\gamma}+(-1)^{1/4}\sqrt{2}\bar{C}^{1/4}_{3}r\right)\\
   &&+c_4 D_{\bar{\upsilon}}\left(i\bar{\gamma}+i(-1)^{1/4}\sqrt{2}\bar{C}^{1/4}_{3}r\right).
\end{eqnarray}

Here, in equation (\ref{d7}), the 'bar' notation includes the parameters  $m$ and $k$, which are replaced with $-m$ and $-k$  in the parameters of equation (\ref{d6}). When examining the energy $E$ of this system, one must solve the equation given below:
\begin{equation}\label{d8}
   \nu=\frac{iC^{2}_{2}-4iC_1 C_3-4C^{3/2}_{3}}{8C^{3/2}_{3}}.
\end{equation}
From the above equation, by setting values to the free parameters, we can numerically obtain the eigenvalues. However, for any configuration of the parameters, the eigenvalues are complex-valued, which is an indication of the absence of bound states. In fact complex-valued eigenvalues of energy, commonly expressed as Quasi-Normal-Modes (QNM), of the massless Dirac equation are present for black holes solutions \cite{Li:2013fka}, therefore is reasonable to expect them for black strings solutions as well. 
\\
\subsection{Solution for $r= \infty$}
For this case, our starting point is the equation (\ref{d1}). If we take $\rho(r)$ as below

\begin{equation}\label{d9}
  \rho(r)=\frac{r^{3}\alpha^{3}+2\mu+2r\alpha A(r)+2r^{2} \alpha A'(r)}{4\alpha r^{2}A(r)}
\end{equation}
and use it in $\psi_1(r)=e^{\int^{r} \rho(r') dr'} Y_1(r)$, we get
\begin{equation}\label{d10}
   -A(r)^{2}Y''_{1}(r)-2A(r)A'(r)Y'_{1}(r)+V_{eff}Y_1(r)=E^{2}Y_1(r)
\end{equation}
where
\begin{eqnarray}\label{d11}
  \nonumber V_{eff}(r)&=&-\frac{1}{4}A'(r) -\frac{(-k+m \alpha)^{2}}{r^{2}\alpha^{2}}+\frac{i(k-m \alpha)A(r)^{3/2}}{r^{2}\alpha} \\ 
  &&+\frac{i(-k+m \alpha)\sqrt{A(r)A'(r)}}{2r \alpha}-\frac{1}{2}A(r)A''(r).
\end{eqnarray}
Expanding $V_{eff}(r)$ in series for $r =\infty$, it becomes 
\begin{equation}\label{d12}
  V_{\infty}(r)= -2\alpha^{4}r^{2}-(k-m\alpha)^{2}+\frac{4\alpha \mu}{r}+\frac{6i(-k+m \alpha)\mu}{\sqrt{\alpha}r^{2}}+O\left(\frac{1}{r}\right)^{3}.
\end{equation}
The system in (\ref{d10}) is not solvable with (\ref{d12}). This time, we apply $\alpha << 1 $ to (\ref{d10}) in order to have simplicity in our calculations, hence, (\ref{d10}) can be obtained as
\begin{equation}\label{d13}
   -\frac{16 \mu^{2}}{\alpha^{2} r^{2}}Y''_{1}(r)+\frac{32 \mu^{2}}{\alpha^{2} r^{3}}Y'_{1}(r)+V_{\infty}(r)Y_1(r)=E^{2}Y_1(r).
\end{equation}
After applying $Y_1(r)=r y(r)$ in (\ref{d13}), and arranging the equation,  we obtain
\begin{equation}\label{14}
    y''+\left(\frac{(E^{2}+k)r^{2}}{16 \mu^{2}}-\frac{2}{r^{2}}+\frac{3ik\alpha^{3/2}}{8\mu}\right)y(r)=0
\end{equation}
where we use $\alpha << 1 $.  Then, $y(r)$ is obtained as
\begin{equation}\label{d15}
   y_{_{n}}=\frac{2^{1/4}}{\sqrt{r}}\exp\left({\frac{i\sqrt{E^{2}+k}\alpha r^{2}}{8\mu}}\right)r^{5/2}L^{3/2}_{n}\left(-\frac{i(E^{2}+k)^{1/2}r^{2}\alpha}{4\mu}\right),
\end{equation}
where $L^{b}_{a}(c r)$ are the Laguerre polynomials. Now we can introduce the energy levels as
\begin{equation}\label{d16}
   E_{n}=\pm \sqrt{k} \frac{\sqrt{9k\alpha-4(5+4n)^{2}}}{2(5+4n)}.
\end{equation}
Here we note that 
\begin{equation}\label{d17}
    9k\alpha>4(5+4n)^{2}
\end{equation}
for the real energy eigenvalues.


\section{Final Remarks}
 
In this paper we have studied the massless Dirac equation in the Black string spacetime. In order to properly take into account the geometric effects on the Dirac's equation we have obtained the tetrads and computed the spin connection. For the massless case, we have a two component spinor in which each component stands for the up and down spin configurations. Due to the symmetries of the system, we have obtained two coupled first-order partial differential equations for the two components of the spinor. These equations could be decoupled, giving rise to two second-order partial differential equations for each component of the spinor. 

The obtained solutions allowed us to understand the behaviour of massless fermions in terms of the cosmological constant, the black string's mass and the radial distance of the particle from the black string. The solutions, written in terms of Parabolic Cylinder functions and Laguerre polynomials, were obtained for a particle far from the black string and around the horizon event. For the particle around the event horizon, for all configuration of parameters, the energy eigenvalues are complex-valued, indicating QNM similarly to the case of spherical black holes. For the particle far from the black string, the energies derived from the Weyl equation set up conditions on the parameters in order to keep the energy as a real valued parameter.

\acknowledgments

\hspace{0.5cm} RD thank the Coordena\c{c}\~{a}o de Aperfei\c{c}oamento de Pessoal de N\'{i}vel Superior (CAPES) for financial support. JF would like to thank the Fundação Cearense de Apoio ao Desenvolvimento Cient\'{i}fico e Tecnol\'{o}gico (FUNCAP) under the grant PRONEM PNE0112-00085.01.00/16 for financial support, the National Council for scientific and technological support (CNPq) under the grant 304485/2023-3 and Gazi University for the kind hospitality.


\appendix

\section{Coefficiets}

The coefficients appearing in the solutions for $r=r_h$ are given by
\begin{widetext}
\begin{eqnarray}
C_1&=&\frac{\alpha ^3 r_h^4 \left(e^2+(k-\alpha  m)^2\right)+12 \alpha ^4 r_h^3 \mu -2 r_h \mu  (k-\alpha  m) \left(-3 i \alpha  \sqrt{r_h^2 \alpha ^2-\frac{4 \mu }{r_h \alpha }}+2 k-2 \alpha  m\right)-12 \alpha  \mu ^2}{\alpha  \left(r_h^4 \alpha ^3-4 r_h \mu \right)^2}\\
\nonumber C_2&=&-\frac{3 i \left(r_h^3 \alpha ^3-2 \mu \right) (\alpha  m-k)}{r_h^3 \alpha  \left(r_h^3 \alpha ^3-4 \mu \right) \sqrt{\frac{r_h^3 \alpha ^3-4 \mu }{r_h \alpha }}}+\frac{2}{r_h^3}+\frac{3 i \left(r_h^6 \alpha ^6+6 r_h^3 \alpha ^3 \mu -4 \mu ^2\right) (\alpha  m-k)}{r_h^3 \alpha  \left(r_h^3 \alpha ^3-4 \mu \right)^2 \sqrt{\frac{r_h^3 \alpha ^3-4 \mu }{r_h \alpha }}}-\frac{4 \left(r_h^3 \alpha ^3-\mu \right) \left(k^2-2 \alpha  k m+\alpha ^2 m^2\right)}{r_h^2 \alpha  \left(r_h^3 \alpha ^3-4 \mu \right)^2}\\
&&-\frac{2 \left(r_h^3 \alpha ^3+2 \mu \right) \left(r_h^6 \alpha ^6+2 r_h^4 \alpha ^2 e^2+16 r_h^3 \alpha ^3 \mu -8 \mu ^2\right)}{\left(r_h^4 \alpha ^3-4 r_h \mu \right)^3}\\
\nonumber C_3&=&-\frac{3}{r_h^4}+\frac{2 \alpha ^2 e^2 \left(5 r_h^6 \alpha ^6+32 r_h^3 \alpha ^3 \mu +8 \mu ^2\right)}{\left(r_h^3 \alpha ^3-4 \mu \right)^4}+\frac{2 \left(5 r_h^6 \alpha ^6-4 r_h^3 \alpha ^3 \mu +8 \mu ^2\right) \left(k^2-2 \alpha  k m+\alpha ^2 m^2\right)}{r_h^3 \alpha  \left(r_h^3 \alpha ^3-4 \mu \right)^3}\\
\nonumber &&+\frac{6 i \left(r_h^6 \alpha ^6-3 r_h^3 \alpha ^3 \mu +5 \mu ^2\right) (\alpha  m-k)}{r_h^4 \alpha  \left(r_h^3 \alpha ^3-4 \mu \right)^2 \sqrt{\frac{r_h^3 \alpha ^3-4 \mu }{r_h \alpha }}}-\frac{6 i \left(r_h^9 \alpha ^9+14 r_h^6 \alpha ^6 \mu -7 r_h^3 \alpha ^3 \mu ^2+10 \mu ^3\right) (\alpha  m-k)}{r_h^4 \alpha  \left(r_h^3 \alpha ^3-4 \mu \right)^3 \sqrt{\frac{r_h^3 \alpha ^3-4 \mu }{r_h \alpha }}}\\
&&+\frac{3 \left(r_h^{12} \alpha ^{12}+44 r_h^9 \alpha ^9 \mu +144 r_h^6 \alpha ^6 \mu ^2-64 r_h^3 \alpha ^3 \mu ^3+64 \mu ^4\right)}{r_h^4 \left(r_h^3 \alpha ^3-4 \mu \right)^4}
\end{eqnarray}
\end{widetext}

\end{document}